\documentstyle[12pt]{article}
\newcommand{\lsim}
{\raisebox{-0.7ex}{$\;\stackrel{\textstyle <}{\textstyle\sim}\;$}}

\twocolumn
\begin{document}
\tiny                       
\begin{center}{
\large 
\bf
ENHANCEMENT OF HIGH-FREQUENCY FIELD IN
NEAR-IDEAL METAL MIXTURE } 
\end{center}
\begin{center}
{\large E. M. Baskin${}^1$, M.V. Entin${}^1$,  A.K. 
Sarychev${}^2$ , A.A.  Snarskii${}^3$}\\
\medskip
{\it ${}^1$Institute of 
Semiconductor Physics, Russian Academy of Sciences, Siberian 
Branch, Prosp.  Lavrentyeva, 13, Novosibirsk, 630090, RUSSIA, 
E-mail:Entin@isp.nsc.ru\\ ${}^2$Scientific Center for Applied Problems in 
Electrodynamics Russian Academy of Sciences, IVTAN, Moscow, 127412, 
RUSSIA,\\ ${}^3$Kiev Polytechnic Institute, pr.Peremoga, 37.  252056, Kiev,  
UKRAINE } 
\end{center}

\begin{abstract} 

The a.c. electric field distribution is studied in a composite, 
containing materials with small local losses.  The  possibility that 
2D two-component system  can have effective absorption in the absence 
of local absorption was evidented \cite{1}. We study the second and 
higher moments of electric field and show that they diverge with the 
decrease of local absorption. The 3D system near the percolation 
threshold is also discussed.  The observation of giant field 
fluctuations in granular films first discovered in \cite{2} is explained.

   \end{abstract}


\subsection*{Introduction} 
Recently \cite{2} the effect
of giant a.c. field fluctuations in granular films with small local 
absorption coefficient was discovered.
This was demonstrated 
by computer simulations on a system, containing discrete elements, 
capacitors and inductors, together with measurements of field 
distribution in a 2D model metal-insulator mixture.

The strong
fluctuations can be explained by the presence of global losses in the
inhomogeneous system, while the coefficients, determining the local 
losses are small \cite{1}, with a consequent local accumulation of 
electromagnetic energy. In the electrotechnical language  local 
energy accumulators are formed by random local $L-C$ circuits or 
using the solid state physics language, by local plasmon modes.

The presence of strong field fluctuations can be interpreted by poor
convergence of effective permeability of medium. This opposes to the
usual convergency of effective conductivity of random system.

Regular
fractals built from hierarchical impedance chain were studied in
\cite{Clerc}  and \cite{EE}. It was demonstrated that in these
systems both impedance \cite{Clerc} and the correlation radius
\cite{EE} are fractal functions of frequency in the lossless 
limit.  Such systems  exhibit the divergency of impedance when 
it size goes to infinity.

The  purpose of the present study is a theoretical explanation of
the discovered effect in the framework of exactly solvable model 
of random media by Dykhne \cite{1} and in a  two-phase near 
ideal metal-insulator 3D model near the percolation threshold.  
We shall show that if the real part of local conductivity
falls down, the effective rate of losses stays finite, resulting 
in the divergency of moments of  electrical field modulus higher 
than second power.  We shall estimate the correlation length on 
a.c.  It is found that the correlation length can infinitely 
grow in a lossless medium.  This corresponds to the situation 
when the system can be transformed to one, contained the only 
local spots, resonating on applied frequency, incorporated into 
moderately inhomogeneous matrix.

\subsection*{Dykhne Medium}
The problem of 2D statistically equivalent two-phase mixture was
first studied by A. Dykhne \cite{1}. The effective dielectric
permittivity $\varepsilon_{\mathrm{eff}}$ in such medium obeys the law
 \begin{equation}
\label{1}
\varepsilon_{\mathrm{eff}} =\sqrt{\varepsilon_{1}\varepsilon_{2}},
\end{equation}
where $\varepsilon_{1}$ and $\varepsilon_{2}$ are the local dielectric
permittivities of both media.

Let's consider the case of two metal media such that on
the field frequency $\omega$ both metals are near-lossless:
 \begin{equation}
\label{2}
\varepsilon_{1,2}
=1-\frac{\omega_{p(1,2)}^2}{\omega(\omega+\frac{i}{\tau_{1,2}})}.
\end{equation}
We shall assume that the field frequency lies
between two plasma frequencies $\omega_{p1}$ and $\omega_{p2}$, say
$\omega_{p1}<\omega<\omega_{p2}$, and $\omega \tau\gg 1$. The 
last inequality tends to the small real part of conductivity 
$\sigma=\frac{\omega}{4\pi
}~\mbox{Im}~\varepsilon$.
 For
bi-metal model $\sigma_{(1,2)}=\frac{\omega_{p(1,2)}^2}{4\pi\omega^2
\tau_{(1,2)}}$.

Such system may be considered, for example, as an infinite lattice,
built from capacitors  and inductors with impedances
$i/\omega C$ and $-i\omega L + r_0$, where $C,L$
and $r_0$ are the capacitance, self-inductance and resistivity of
elements, mixed in equal ratio.

Within the region $\omega_{p1}<\omega<\omega_{p2}$ both
permittivities $\epsilon_{1,2}$ are real, so lossless, while
$\epsilon_{\mathrm{eff}}$ is
imaginary.  Nevertheless if the local rate of absorption is small,
the system as a whole absorbs energy.

The sign of $\varepsilon_{\mathrm{eff}}$
should be chosen positive in accordance with the condition of energy
absorption. The imaginary conflict with the energy conservation law is
resolved similar to Landau damping --- the absorption of the whole media
means the excitation of local plasmons. That gives losses of a macroscopic
medium while the microscopic media are lossless. Hence the energy
should be absorbed by the whole medium without heating and should be
accumulated by plasmons up to it will convert to the heat by real
conductivity.

In the case of Dykhne medium we can find the exact formulae for 
$\langle |{\bf E}|^2 \rangle$  and $\langle {\bf E}^2 \rangle $.
From the energy conservation law we have
 \begin{equation}
\label{3}
\sigma_{\mathrm{eff}}|\langle{\bf E}\rangle|^2=\langle\sigma |{\bf
E}|^2\rangle,
\end{equation}
where $\sigma_{\mathrm{eff}}=\frac{\omega}{4\pi }~\mbox{Im}~\varepsilon_{\mathrm{eff}}$. 

Dykhne \cite{1} obtained the equations for mean even powers of
field modules in different phases of bi-metal medium
\begin{equation} \label{31}
\sigma_1^{n} \langle|{\bf E}|^{2n}\rangle_1  =  \sigma_2^{n}
\langle|{\bf E}|^{2n}\rangle_2 \end{equation}
(The subscripts 1 and 2 means that averaging is done in the
first or second media correspondingly).

The equation (\ref{31}) may be generalized to the a.c. case 
including complex conjugation and vector components of field 
(Appendix):

\begin{equation} \label{32}
\varepsilon_1^{n/2} (\varepsilon_1^*)^{m/2} \langle \overbrace{{\bf E}_i {\bf
E}_j \ldots {\bf E}_k}^n \overbrace{{\bf E}_r^* {\bf E}_p^* \ldots {\bf
E}_q^*}^m \rangle_1 = \varepsilon_2^{n/2} (\varepsilon_2^*)^{m/2}\langle {\bf
E} _i \ldots{\bf {E}^*}_q \rangle _2. \end{equation}
Combining (\ref{3}) and  (\ref{32}) for $m=1, n=1$ or $m=2, n=0$, we obtain
\begin{eqnarray}\label{34}
\langle |{\bf E}|^2 \rangle
&=&\frac{(1+|\varepsilon_2/\varepsilon_1|) \sigma_{\mathrm{eff}} }{\sigma_1
|\varepsilon_2/\varepsilon_1|+\sigma_2} |\langle {\bf E}\rangle|^2\\
\langle {\bf E}^2 \rangle &=&\frac{1}{2}
\varepsilon_{\mathrm{eff}}\frac{\varepsilon_1+\varepsilon_2}{\varepsilon_1
\varepsilon_2}\langle {\bf E}\rangle^2. \end{eqnarray}
The equations (\ref{3}) and (\ref{34}) show that  $\langle |{\bf
E}|^2 \rangle $ diverges in a weakly absorbing medium like reciprocal
absorption, while  $\langle {\bf E}^2
\rangle $ remains finite.
 This divergency
does not contradict to the finite value of average field due 
to the phase shift of field in different parts of system.

The divergency of higher moments  follows from the divergency of
second moment and the Cauchy-Bunakovskii inequality
\begin{eqnarray}\label{35}
\langle |{\bf E}|^{2n}
\rangle \geq \langle |{\bf E}|^{2} \rangle ^n.
\end{eqnarray}
The equation (\ref{32}) give the phase
shifts between mean powers of fields in different media. 
In  particular, the mean fields 
in the first and the second media have  phase shift $\pi/2$.  

\subsection*{Correlation length of electric field}
 The application of high frequency field leads to 
appearance of "hot" spots, resonantly accumulating the 
 energy. The spatial distribution of them is random and depends 
 on the frequency. From that point of view  any medium 
 containing regions with both positive and negative 
permittivities can be transformed to  rare gas of hot spots, 
incorporated into a weakly inhomogeneous  matrix.

In the framework of L-C model  the electric circuit has
one resonant frequency per a cell.  All of them are distributed
within the range $\omega_{p2}- \omega_{p1}$. The typical width of
resonance is determined by $\sigma$. Thus the density of resonances
per unite frequency interval is $ \sigma/(\omega_{p2}- 
\omega_{p1})$ and the mean distance between them grows like 
\begin{equation} \label{33} ((\omega_{p2}- 
\omega_{p1})/\sigma)^{1/2}.  \end{equation} This is a specific 
correlation length of electric field.

From the other hand, we can use the "hot spots" language. Let's
change the system to a 2D system, containing round
inclusions with radius $a$ and permittivity $\varepsilon_2$ into 
a matrix with permittivity $\varepsilon_1$.  The low density 
approximation give 
\begin{eqnarray} \label{38} {\bf 
E}=\langle{\bf E}\rangle + \sum_n{\frac{{\bf p_n(r-r}_n)}{({\bf 
r-r}_n)^2}}\\ {\bf p_n}=\chi {\bf E}({\bf r}_n), \end{eqnarray} 
where $\chi= \frac{\varepsilon_1- 
\varepsilon_2}{\varepsilon_1+\varepsilon_2}\frac{a^2}{2}$ 
is the polarizability of inclusions.  If $\chi$ is not too large we
can substitute $\langle{\bf E}\rangle$ instead of ${\bf E}({\bf
r}_n)$. The effective permittivity in this model is

\begin{equation} \label{39}
\varepsilon_{\mathrm{eff}}=\varepsilon_1(1+4\pi n\chi),
\end{equation}
where $n$ is the density  of  inclusions. The polarizability of
inclusions diverges if $\varepsilon_1+\varepsilon_2=0$, which corresponds
to a local plasmon mode.

We can estimate the correlator of potential
fluctuations $\delta \varphi ({\bf r})$ as

\begin{equation} \label{40}
\langle(\delta \varphi (0){\bf\delta \varphi ( r})\rangle=n\int
\frac{({\bf p(r-r'))}}{\bf (r-r')^2}\frac{({\bf pr')}}{\bf r'^2}d{\bf
r'}\sim\pi n (\chi {\bf \langle E\rangle})^2\log
\frac{R}{r}.\end{equation}
$R$ is a cutoff radius, included to limit
the divergency of integral in (\ref{40}).  The correlation length can
be estimated from (\ref{40}) as a length scale, where the 
fluctuating potential is comparable with the potential of 
average field $\langle{\bf E}\rangle L_c$:

  \begin{equation} \label{41}
  L_c\sim (n\chi^2)^{1/2}\log \frac{n\chi^2}{a}.
\end{equation}
Really if the field fluctuations  exceed the average
field $\langle{\bf E}\rangle$  we cannot consider the
polarization ${\bf p}$ as a preassigned quantity and should change
it together with the electric field. Hence $L_c$ plays role of 
limiting size for the formula (\ref{40}) and $R=L_c$.

The density of inclusions $n$ in Dykhne medium can be estimated as
a density of "hot spots"
$n=\frac{\sigma}{\omega_{p2}-\omega_{p1}}$. The substitution of 
this value to (\ref{41}) give the formula  (\ref{33}) with the 
accuracy of logarithm.

From the formula (\ref{34}) it follows that
$\langle |{\bf E}|^2\rangle$  grows like  $1/\sigma$
and the distance between hot spots grows proportionally
$\sigma^{-1/2}$.  This shows a good agreement with the experimental
results and computer simulations  \cite{2}.

The ultimate "hot spots" language  does not correspond to the
power-law correlation of electric field. Nevertheless it permits to
estimate the correlation length in the a.c. problem.

The question about correlation length
of electric field is not so simple. The correlation length in the
problem of percolation is a geometric property, limiting the power
behavior of probability of two distant sites to belong to the same
cluster. It is independent on electric parameters.

From the other hand in the static conductivity (or dielectric
response) problem the correlation length in the smearing domain is
determined by the spatial size where the  opportunity of current to
flow in both media is equal and both phases give the same
contributions to the effective conductivity (dielectric constant).

If both media have positive
permittivities or the imaginary part of permittivities is comparable
with real part, the system is not resonating, and the correlation
radius conserves the same order of magnitude as in a stationary case.
In the resonant case without local losses the correlation length is
infinite.

In such system the response should depend on a specific
realization of medium. The spatial distribution of field becomes
fractal and the electric properties are not self-averaging. For
example, this is reflecting in the frequency dependence of
correlation radius \cite{EE}. The system spectrum contains 
separate lines. This is easy to understand if to consider 
the lossless system like the quantum particle (photon) 
propagation in the inhomogeneous disordered medium. In fact, the 
spectrum of quantum system consists from   distinct lines and 
the spatial distribution of wave function amplitude is fractal. 
The divergency of squared module of electric field corresponds 
to the localization of photons. In the terms of L-C model this 
is the localization of field in resonating circuits.

The finite real part of conductivity limits the divergency of field
and the correlation radius but retains the complicated structure of
field.

  \subsection*{Higher dimensions}
Unlike 2D case there is no exact solution for
effective coefficients in the 3D system, but the percolation theory gives
the effective constants for the  strongly inhomogeneous medium limit.

If the concentration of the binary mixture is near the percolation
threshold, while both permittivities are finite, the effective
permittivity does not depend on the concentration $\tau$ but on the
ratio of permittivities only. This is so called smearing domain
\cite{4}. If additionally the ratio of permittivities is small,
$\frac{\varepsilon_1}{\varepsilon_2}=h\ll 1$, a power 
dependence is valid 
\begin{equation} \label{6} 
 \varepsilon_{\mathrm{eff}}=\varepsilon_2 h^{\alpha}. \end{equation} 
The value of $h$ is small,  if $\omega_{p1}\ll \omega_{p2}$ and 
$\omega_{p1}<\omega < \omega_{p2}$.

Above the percolation threshold ($\tau>0$) the width of smearing
domain is $\tau\lsim h^{\alpha/t}$.

These estimations are usually applied to describe the system with
positive conductivities (dielectric constants).
The analyticity permits to continue them for the negative signs 
of constants too. The equation (\ref{6}) tends to the complexity 
of effective permittivity in the mixture of domains with 
positive and negative permittivities. The phase of 
$\varepsilon_{\mathrm{eff}}$ is universal and is determined by a critical 
exponent only:  
\begin{equation} \label{7} Arg~ 
\varepsilon_{\mathrm{eff}}=(2n+1)\pi \alpha +\pi. \end{equation} 
Analogously Dykhne medium,  the root branch  $n=0$
can  be chosen if one account for  the absence of permittivity
discontinuities in the complex plane except for the negative part
of abscissa.

Like 2D, the small
local losses may be accompanied by the finite effective
rate of absorption.  Instead of the equation (\ref{3}) we can use
the inequalities for the mean square of field.  Using \cite{3}, we
obtain the inequalities

\begin{equation} \label{30}
\langle\frac{1}{\sigma}\rangle\sigma_{\mathrm{eff}}|\langle{\bf
E}\rangle|^2>\langle|{\bf E}|^2\rangle
>\frac{\sigma_{\mathrm{eff}}}{\langle\sigma\rangle }|\langle {\bf
E}\rangle |^2. \end{equation}

The formula (\ref{30}) shows that
mean square of field also diverges. The rate of divergency can be
estimated if we assume that both media have the same  conductivities
$\sigma ~=~\sigma_1~=~\sigma_2~$, when the inequalities convert to
the equation
\begin{equation} \label{4} \langle |{\bf E}|^2\rangle =
\frac{\sigma_{\mathrm{eff}}}{\sigma}|\langle {\bf E}\rangle |^2. \end{equation}
The last equation coincides with the consequence of Eq. \ref{3} in 2D case
 with the same assumptions.
\subsection*{Conclusions}
 In the present work we studied  the  moments of 
electric field in the Dykhne media with low absorption. We 
exactly found the second moments and showed that $\langle\sigma |{\bf
E}|^2\rangle$ diverges and $\langle\sigma {\bf
E}^2\rangle$  is limited if the absorption goes down. We 
couldn't find higher moments but proved the relation between 
mean values of $|{\bf E}|^2$ in two materials and the divergency 
of the higher moments  $\langle |{\bf E}|^{2n}\rangle$.

The divergency is conditioned by the different signs of media 
permittivities responsible for the existance of local plasmons. 
The divergency of field moments is not specific for 
2D 2 -components composite only. The same result is obtained for 
3D system in the smearing domain.

The divergency of higher moments leads to the  divergency of 
field correlation length in the system with low local 
absorption.  Usually the system is described by the effective 
permittivity or the effective conductivity which have small 
fluctuations in a large system. This thermodynamic limit 
requires smallness of the correlation length compared to the 
system size and hence finite absorption. If this length exceeds 
the size of the system the effective properties do not have 
definite limits and they have fractal dependencies on system 
size. This is the behavior evidenced in regular fractal circuits 
\cite{EE}.

We should underline the difference of static and a.c. cases. In 
the static case the correlation length is determined by the 
ratio of dielectric constants or conductivities of media. In the 
high-frequency case the correlation length is not connected with 
the absolute values of permittivities but with the small 
conductivity, determining the absorption. If we consider the 
case of ideal conductors or insulators with zero absorption, the 
static effective dielectric constant and conductivity are 
definite quantities, while in the a.c. case they are not. 

This conclusion becomes almost trivial if we consider the static 
properties of mixture with positive and negative conductors, 
being formally equivalent to the metal mixture with 
alternating signs of permittivities. The mixture with negative 
conductors  is thermodynamically unstable. This means the 
generation of a.c. and no thermodynamic limit. 

\subsection*{Acknowledgments}
The work was partially supported by Russian Foundation for Basic
Researches (Grants 950204432 and 960219353) and by 
Volkswagen-Stiftung.

\subsection*{Appendix. The relation between field moments.}
To obtain the relation (\ref{32}) we can do Dykhne transform with the
field and the electric displacement ${\bf D}=\varepsilon {\bf E}$ 
\begin{eqnarray} \label{50}
{\bf E'}=1/\varepsilon_{\mathrm{eff}}[{\bf nD}],&~~~~~~~& {\bf 
D'}=\varepsilon_{\mathrm{eff}}[{\bf nE}],\\ \nonumber 
x\to y,& ~~~~~~~&y\to -x. 
\end{eqnarray}
where $\varepsilon=\varepsilon_{1,2}$ in the first and the second medium
correspondingly. It conserves the equations for field and displacement
\begin{eqnarray} \label{51}
\nabla{\bf D~}=0&~~~~~~~~&\nabla\times{\bf E}=0,\\
\nonumber
\nabla{\bf D'}=0&~~~~~~~~&\nabla\times{\bf E'}=0
\end{eqnarray}
and transforms the expression for electric displacement ${\bf
D}=\varepsilon {\bf E}$ to ${\bf
D}'=\frac{\varepsilon}{\varepsilon_{\mathrm{eff}}} {\bf E'}$. Using the
statistical equivalence of both media, we  obtain (\ref{32}).

\end{document}